\documentstyle[12pt,aaspp4]{article}
\def\sec{\hbox{$''$\hskip-2pt.}}
\def\deg{\hbox{$^{\circ}$\hskip-2pt.}}
\begin{document}

\title{DIRECT OBSERVATION OF THE FOURTH STAR IN THE ZETA CANCRI SYSTEM}

\author{J.B. Hutchings\altaffilmark{1}}
 \affil{Herzberg Insitute of Astrophysics,
National Research Council of Canada,\\ Victoria, B.C. V8X 4M6, Canada}

\author{R.F.Griffin}
\affil{The Observatories, Madingley Road,\\
Cambridge CB3 0HA, England}

\author{F. M\'enard\altaffilmark{1}}
\affil{Canada France Hawaii Telescope Corporation,\\
P.O.Box 1597, Kamuela, Hawaii 96743}

\altaffiltext{1}{Observer with the Canada France Hawaii Telescope which
is funded by NRC of Canada, CNRS of France, and the University of Hawaii}

\begin{abstract}

  Direct imaging of the $\zeta$ Cnc system has resolved the fourth star in
the system, which is in orbit around $\zeta$ Cnc C. The presence of the
fourth star has been inferred for many years from irregularities in the
motion of star C, and recently from C's spectroscopic orbit.  However, its
mass is close to that of C, making its non-detection puzzling.  Observing
at wavelengths of 1.2, 1.7, and 2.2 $\mu$ with the adaptive-optics system
of the CFHT, we have obtained images which very clearly reveal star D and
show it to have the color of an M2 star. Its brightness is consonant with
its being {\it two\/} M stars, which are not resolved in our observations
but are likely to be in a short-period orbit, thereby accounting for the
large mass and the difficulty of detection at optical wavelengths, where
the magnitude difference is much larger. The positions and colors of all
four stars in the system are reported and are consistent with the most
recent astrometric observations.

\end{abstract}


\section{Introduction}

   $\zeta$ Cancri is a visual triple stellar system which has been studied
as such for over 200 years.  The system consists of three bright stars, all
dwarfs of close to solar type with visible magnitudes near 6. The orbital
planes of the system are not far from the plane of the sky.  There is a
close pair AB separated by about 1 arcsec with an orbital period 58.9
years, and a third star C about 6 arcsec away with an orbital period
estimated at 1100 years. Star C itself has proper-motion irregularities
that were first remarked upon by Sir John Herschel (1831) and were
interpreted by Otto Struve (1874) in terms of orbital motion in a period on
the order of 20 years around an unseen companion.  Subsequent observations
and discussions have corroborated Struve's interpretation, and have
recently been reinforced by a spectroscopic orbit of C by Griffin (2000),
who has also described in some detail the interesting observational history
of the system.  The situation concerning the three visual orbits in which
the $\zeta$ Cnc stars are involved (those of AB, CD, and AB--CD) have been
summarized most recently by Heintz (1996).  All three have appreciable
eccentricity, with that of CD being lowest at 0.12.

   The principal question that has remained outstanding concerning the
$\zeta$ Cnc system is the nature of the companion to star C, since its
orbit indicates a companion of comparable mass and a separation of about
0.4 arcsec.  Griffin (2000) recounts some of the earlier claims regarding
the nature of star D, and it is clear that it must be considerably fainter
than C at visible wavelengths. Suggestions have consequently been made that
it is a white dwarf, or else a close pair of M stars which would be quite
faint in the visible.  Griffin concluded that the question remained open,
but that a short exposure by a large telescope with adaptive optics might
provide the definitive evidence. Accordingly, the system was observed
during a run with the Canada France Hawaii telescope by one of us (JBH) in
January 2000, and followed up by further data from the same system a month
later by another of us (FM). As we describe below, we discovered star D
immediately and easily.

\section{Observations and data}

    The observations of $\zeta$ Cnc were carried out during runs of the
adaptive optics (AO) camera of the Canada France Hawaii telescope (CFHT),
with the near-infrared camera KIR (see Rigaut et al 1998). 
They were obtained during gaps in other
programs, using filters that happened to be in the camera. Thus, as
described below, the observing program was not particularly well planned,
but it occupied a total exposure time of under one minute! The field of view
of 35$'$ easily includes all of the stars in the system.  The brightness of
the stars made the AO correction extremely good, with FWHM close to the
diffraction limit at all wavelengths. The filters used were generally
narrow-band, to avoid detector saturation and the need for very short
exposures. Table 1 lists the details of the observations.

   The first observations by one of us (JBH) in January 2000 used the AB
pair as guide star. That proved not to be a happy choice: the two bright
stars are separated by less than 1$''$ and the guiding system \lq hunted'
back and forth between the two, producing double images of all the stars
present, all with attendant diffraction rings!  The `PSF' pattern also
changed depending on where in the field of view the guide star(s) were,
during the dither pattern used. In spite of the complex images produced, it
was clear that a companion star to C (star D) was present, and it was
possible to measure its position and brightness by subtracting the complex
PSF. The process was fortuitously made simpler by the fact that star D is
quite well separated from C along the same direction as the separation of
AB (and hence the double PSF).

   Since it was clear that better results would be obtained by guiding on
star C, which is effectively single at the visible wavelengths used for
guiding, and in view of the possibility of measuring the relative motions
of the stars over an interval of several weeks, further observations were
obtained in February by one of us (FM). The AO system produced very 
well-corrected images from the bright guide star, with FWHM of 0\sec 08, 
0\sec 10, and 0\sec 12 at $J$, $H$, and $K$; they are shown in Figure 1.

   Since the stars are so bright and the exposures so short, there was no
need for flat fields or sky subtraction in order to measure the star fluxes
and positions. Small changes seen in the diffraction patterns are due to
the shortness of the exposure times. The residual seeing halo due to any
incompleteness of the AO correction is resolved into individual speckles,
as is typical for short exposures on AO systems. However, we established
that measurements from images dithered to different places on the detector
produced values that differed by amounts too small to matter in this
investigation.

\section{Measurements and discussion}

   For each of the observing runs, the positions and fluxes of the stars
were determined, using all the images taken. Fluxes were measured by the
IRAF task `imexam', and also by summing the signal from sections of image
and subtraction of the off-star signal levels. Star positions were determined
by imexam, and also by visual inspection of contour plots over the central
pixels of each star image. 

   The camera was removed and re-installed between the two observing runs.
Thus, it was necessary to check the orientation of the images for small
rotations. Since systems AB and CD moved in their orbits over the interval,
we used the position angle of BC as the fiducial, since its predicted
change in the interval is only 0\deg 03 (Heintz 1996). That check showed 
that a marginal rotation
between runs of 0\deg 05$\pm$0\deg08 had occurred. This correction has
been applied to the position-angle differences we report.

   Our main interest is in relative fluxes between the system stars, so no
absolute calibration was applied, except to check that the overall colors
between visible and IR are as expected for the bright stars, which are all of
almost the same spectral type. Table 2 shows the measurements made of fluxes
and Table 3 shows the position and angle measurements.

    Table 2 compares the color differences expected for main-sequence
stars of various likely types, from Bessell \& Brett (1988).  They have been
corrected from the standard $JHK$ values to account for the different
bandpass centres of the Fe II and Br$\gamma$ filters. The spectral type of
star C is quoted as G5 in the Bright Star Catalogue (Hoffleit 1982) but
Griffin (2000) has argued that it is based upon a misconception and that G0
is more likely. As A and B are bright and overlapping, while C and D have
different fluxes, the color differences C$-$D are more reliable than A$-$D
or B$-$D. 

   As a figure of merit, we sum the differences between model and observed
colors in Table 2. For B$-$D, that indicates that M2 is better than M0 for
D, and for C$-$D G0+M2 fits better than G0+M0 or G5+M2. Thus, we conclude
from the color differences that C and D have spectral types G0 and M2,
respectively, with an uncertainty only on the order of one spectral
subtype.

     The mass ratio C/D is close to unity --- Heintz (1996) gives masses of
0.99 and 0.93 for C and D --- which does not accord with the mass ratio of
single main-sequence stars of types G and M. However, if D is itself a
binary system of two stars close to spectral type M2, the mass would be
0.78, and for an M0 pair it would be 0.94 (see Allen 1973).  From the mass
function of 0.05 found by both Griffin and Heintz, such a combined mass
corresponds to an orbital inclination of about 35\deg

     The angle changes noted in Table 3 are fully consistent with the
published orbits of Heintz (1996). Note that the orientation of our
detector is based on adopting the predicted angle of BC from Heintz's
orbits. The `predictions' in the table are
derived from Heintz's tables and periods, assuming a mass ratio
near unity for CD. The observed separation and direction of D from C, and
their changes, indicate that to be the case. Monitoring of the CD visual
orbit with further AO images will enable much better estimates to be made
of the mass ratio.

   Given that the stars are all at the same distance, absolute magnitudes
can be used to predict the magnitude difference between C and D in the $V$
band as 4.5 for two M2 stars and 3.5 for two M0 stars, compared with G0
(Allen 1973).  Applying the standard ($V-K$) colors to the difference we
measure between stars C and D we derive $V$ magnitude differences of 4.1
and 3.5 for pairs of M2 and M0 stars, respectively.  Thus, the colors and
relative fluxes are fully consistent with stars D being an unresolved
binary of two M stars.  A challenge for future observations will be to
resolve D at visual wavelengths and measure its $V$ magnitude, and even to
demonstrate (directly and/or spectroscopically) that it is double.

     We are grateful to David Crampton for suggesting the
observations to the CFHT observer after a conversation following 
a colloquium on $\zeta$Cnc given by one of us (RFG).

\vskip40pt

\centerline{References}

Allen C.W., 1973, Astrophysical Quantities, Athlone Press, London (third
edition).

Bessell M.S., \& Brett J.M., 1988, PASP, 100, 1134 

Griffin R.F., 2000, Observatory, 120, 1

Heintz W.D., 1996, AJ, 111, 408

Herschel J.F.W., 1831, Mem.RAS, 5, 171

Hoffleit D., 1982, The Bright Star Catalogue, Yale University Observatory

Rigaut F., et al, 1998, PASP, 110, 152

Struve O., 1874, Comptes Rendus, 79, 1463

\begin{deluxetable}{lllll} 
\tablenum{1}
\tablecaption{$\zeta$ Cnc journal of observations}
\tablehead{\colhead{JD 245..} &\colhead{Filter} 
&\colhead{$\lambda, \Delta\lambda$($\mu$)} &\colhead{Exp (sec)}
&\colhead{Comment} }
\startdata
51568.4 &Br$\gamma$ &2.16, 0.02 &5.0 &Guide on AB, 4 dither\nl
51568.4 &J cont &1.21, 0.015 &5.0,2.0 &Guide on AB, 5 dither\nl
51596.2 &J &1.25, 0.16 &0.5 &10 frames, guide on C\nl
51596.2 &Br$\gamma$ &2.16, 0.02 &1.0 &6 frames, guide on C\nl
51596.2 &FeII cont &1.69, 0.02 &0.7 &5 frames, guide on C\nl
\enddata
\end{deluxetable}

\begin{deluxetable}{llll} 
\tablenum{2}
\tablecaption{$\zeta$ Cnc color differences of stars}
\tablehead{\colhead{Stars} &\colhead{J--FeII} &\colhead{J--Br$\gamma$} 
&\colhead{FeII--Br$\gamma$} }
\startdata
A--B &0.00 &0.00 &0.00\nl
B--D &0.34 &0.47 &0.21\nl
A--C &0.00 &0.03 &0.03\nl
C--D &0.32 &0.51 &0.19\nl
\nl
F9 V -- M2 V &0.31 &0.55 &0.19\nl
G0 V -- M2 V &0.31 &0.53 &0.18\nl
G5 V -- M2 V &0.26 &0.47 &0.18\nl
G0 V -- M0 V &0.23 &0.47 &0.15\nl
\enddata
Color differences $\pm$0.05
\end{deluxetable}

\begin{deluxetable}{lrlll} 
\tablenum{3}
\tablecaption{$\zeta$ Cnc astrometry}
\tablehead{\colhead{Date/change} &\colhead{CD($''$)} &\colhead{CD($^{\circ}$)}
&\colhead{AB($''$)} &\colhead{AB($^{\circ}$)} }
\startdata
51568.4 &0.336$\pm$0.006 &86.2$\pm$1.6 &0.837$\pm$0.003 &83.69$\pm$0.35\nl
51596.2 &0.317$\pm$0.003 &84.5$\pm$0.9 &0.841$\pm$0.004 &82.97$\pm$0.15\nl
Heintz 51596.2 &&&0.835 &83.10\nl
\nl
Change  &$-$0.019$\pm$0.007 &1.7$\pm$1.8 &0.004$\pm$0.005 &0.72$\pm$0.38\nl
Orbit prediction   &--- &1.6 &0.003 &0.42\nl
\enddata
Position angles are as in Heintz (1996) --- from N towards E
\end{deluxetable}

\vskip40pt

\centerline{Caption to Figure 1}

Images of $\zeta$ Cnc guided on star C. N is 2\deg 8 left of vertical, and
E to the left.
Filters are $J$ (top left, 1.2$\mu$), Fe II cont (top right, 1.7$\mu$),
Br$\gamma$ (lower left, 2.2$\mu$). Lower right is detail of CD pair
with Fe II cont filter. CD separation is 0\sec 32.

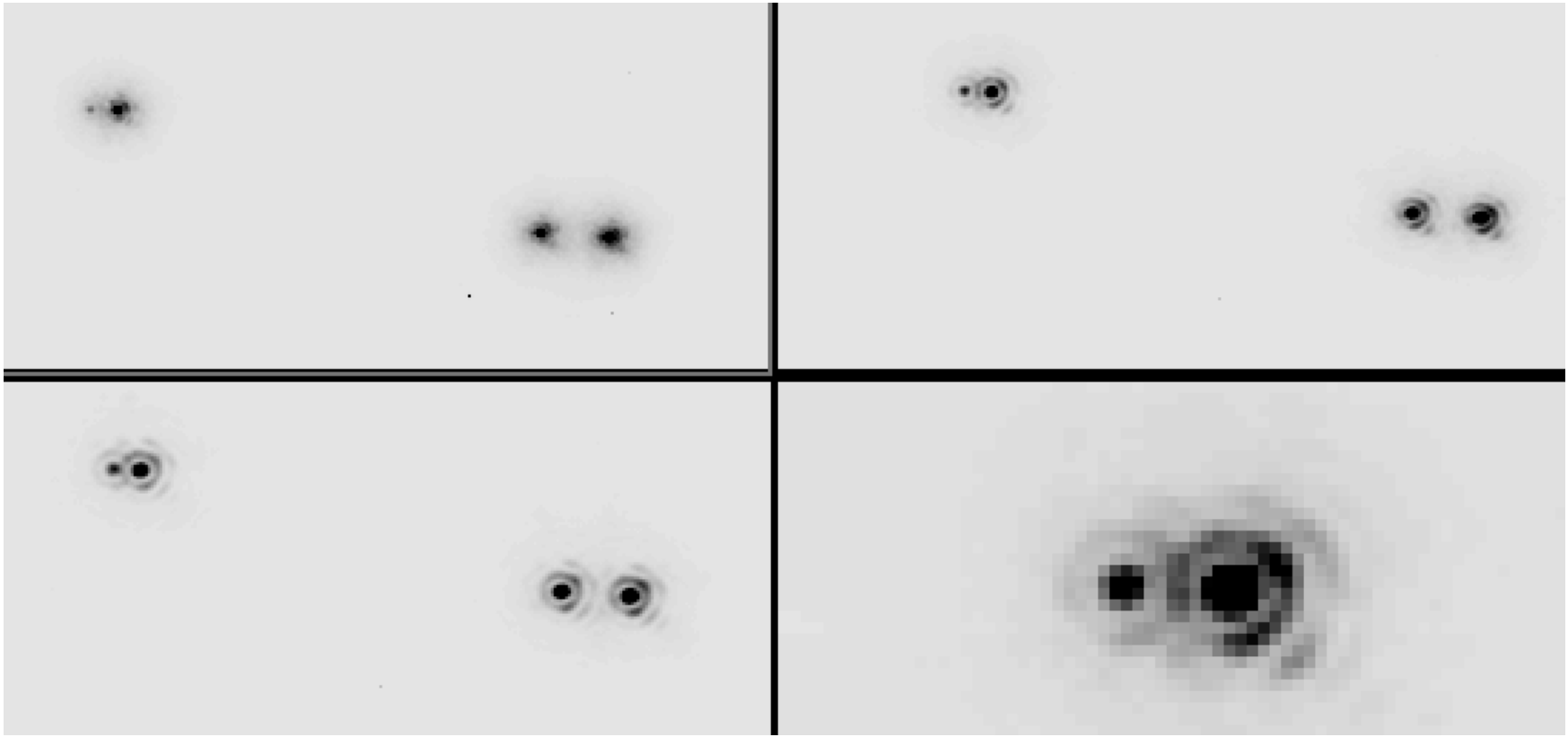
\end{document}